\documentclass[aps,prb,twocolumn,groupedaddress]{revtex4}
\usepackage{graphicx}

\begin{document}

\title{Shot-noise characteristics of triple-barrier resonant-tunneling diodes}
\author{A.K.M. Newaz}
\author{W. Song}
\affiliation{Department of Physics and Astronomy, State University of New York at Stony Brook, Stony Brook, NY 11794-3800}
\author{Y. Lin}
\altaffiliation[Present Address:\,\,\,]{Department of Physics, National Tsing Hua University, Taiwan, Republic of China.}
\author{J. Nitta}
\affiliation{NTT Basic Research Laboratories, Atsugi, Kanagawa, 243-0198, and CREST-JST, Japan}
\author{E. E. Mendez}
\email[e-mail address:\,\,\,]{emendez@notes.cc.sunysb.edu}
\affiliation{Department of Physics and Astronomy, State University of New York at Stony Brook, Stony Brook, NY 11794-3800}

\date{\today}

\begin{abstract}

We have found experimentally that the shot noise in InAlAs-InGaAs-InAlAs Triple-Barrier Resonant-Tunneling Diodes (TBRTD) is reduced over the 2$eI$ Poissonian value whenever their differential conductance is positive, and is enhanced over 2$eI$ when the differential conductance is negative. This behavior, although qualitatively similar to that found in double-barrier diodes, differs from it in important details. In TBRTDs the noise reduction is considerably larger than predicted by a semi-classical model, and the enhancement does not correlate with the strength of the negative differential conductance. These results suggest an incomplete understanding of the noise properties of multiple-barrier heterostructures. 

\end{abstract}

\maketitle


The measurement of shot noise, in combination with electrical conductance, is an important tool to elucidate electronic transport in mesoscopic devices. If the electronic noise is created randomly, its spectral density, $S_I(\omega)$, has the value $2eI$ and we speak of Poissonian or full shot noise. But if the motion of the charged carriers is correlated, then there are deviations (either reduction or enhancement) from the Poissonian value.\cite{Blanter00} A measure of these deviations is given by the so-called Fano factor, $F$, defined as the ratio of the actual noise spectral density to the full shot noise value.

One of the mesoscopic devices that best illustrates non-Poissonian noise due to electron correlation is the \, Double-Barrier \, Resonant-Tunneling Diode \, (DBRTD).\cite{Blanter00} Its current-voltage characteristic ($I$-$V$) usually has a quasi-triangular shape, with an initial region of positive differential conductance (PDC) followed by a sharp negative differential conductance (NDC). Shot noise in a DBRTD is partially suppressed (i.e., it is sub-Poissonian) in the PDC region\cite{Li90} and enhanced above $2eI$, (i.e., super-Poissonian) in the NDC region.\cite{Iannaccone98, Kuznetsov98} The reduction of noise has been explained by correlation effects due to Pauli's exclusion principle,\cite{Li90} while the enhancement has been accounted for by a positive-feedback correlation.\cite{Iannaccone98,Aleshkin2001,Korothkov} The transition from the sub-Poissonian to the super-Poissonian regime at, or near, the current peak has been studied by considering the potential fluctuations induced by charge fluctuations in the quantum well.\cite{Blanter99} Experimentally, it has been shown that the larger the absolute value of the negative differential conductance, the larger the noise enhancement,\cite{Kuznetsov98} and it has been unequivocally established that charge accumulation is essential to the enhancement of shot noise in a DBRTD.\cite{Song2003}

\begin{figure}[thp!]
  \centering
  \includegraphics[angle=0,clip=]{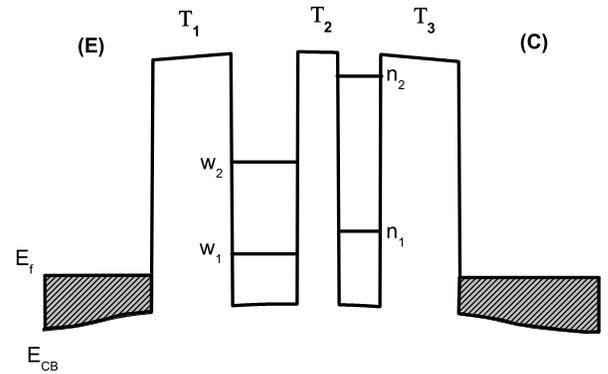}
  \caption[Potential Profile of a Triple Barrier Resonant Tunneling Diode]{\label{fig:Conduction}  Conduction-band profile of sample A (see text) under flat-band condition (zero bias). The solid lines in the quantum wells show the quasi-bound energy levels. The emitter (adjacent to the substrate) and the collector are $n$-type doped regions denoted by {\bf{(E)}} and {\bf{(C)}}, respectively. $T_i$ represents the transmission probability for the $i$-th barrier.}
\end{figure}

\begin{figure}[thp!]
  \centering
  \includegraphics[width=3.2in,angle=0,clip=]{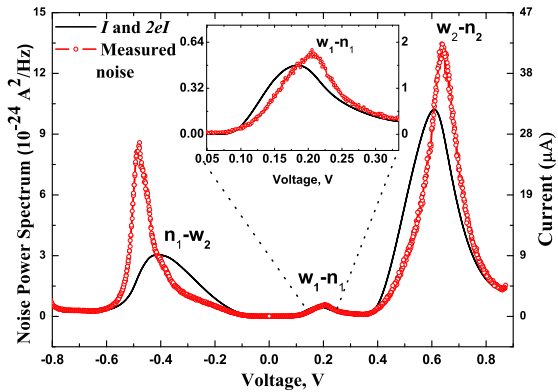}
  \caption[Shot Noise of a Barrier Resonant Tunneling Diode]{\label{fig:noise} Current (solid line) and shot noise (circles) characteristics of sample A (see text), measured with the diode at $T = 4.2$ K. The scales for current and noise, shown on the right and left vertical axes, respectively, differ by a factor of 2$e$ so that the current $I$, read using the left scale, can be regarded as 2$eI$. The inset shows a blow-up of 2$eI$ and the measured shot noise around the $w_1-n_1$ resonance.}
\end{figure}

Shot-noise measurements in DBRTDs have been limited to devices with relatively thick barriers, in which tunneling is sequential. Although several calculations have shown that in multiple-barrier structures the shot-noise reduction should be independent of whether the electronic transport is sequential or coherent,\cite{DeJong95} other calculations have predicted a smaller shot noise when the process is coherent. For instance, while a sequential-tunneling model predicts a minimum sub-Poissonian noise of 0.41 in a triple-barrier structure,\cite{DeJong95} there are calculations that, assuming coherent transport, anticipate a minimum sub-Poissonian Fano factor that ranges from 0.41 to 0.22, depending on the details of the heterostructure.\cite{Sun99,Pouyet03,Pra97} A Triple Barrier Resonant Tunneling Diode (TBRTD) is in principle more suitable than a DBRTD to study the effect of coherence on noise since in the TBRTD the coupling between its two wells can be varied by adjusting the thickness of the middle barrier, while maintaining the low current necessary for noise measurements with thick end barriers.

Before addressing the question of coherence it is important to compare systematically the sequential-tunneling noise of DBRTDs and TBRTDs, especially in light of the very limited experimental information on the latter. In the only study we know, it was found that at the onset of the tunneling current shot noise was $2eI$ and then became progressively smaller, with a minimum value of 0.7 as the current approached its peak value.\cite{Yau97} Such a decrease is surprising, and is at odds with what is predicted theoretically and with the behavior found in DBRTDs.\cite{Blanter00}

As a first step toward the goal of measuring the shot noise of strongly coupled quantum wells, we have studied the noise of thick TBRTDs in both the PDC and NDC regions and compared it with that of a ``control'' DBRTD. We have found that in the PDC region the noise reduction was considerably larger in the TBRTDs than previously observed and theoretically predicted by a semiclassical model. In the NDC region of these devices, we found noise enhancement, as in a DBRTD, but for one of the two bias polarities that enhancement was anomalously large relative to what is found in a DBRTD of comparable negative differential conductance. 

Our TBRTDs and DBRTD were prepared using lattice-matched InGaAs-InAlAs epitaxial layers grown by metal-organic chemical vapor deposition on InP substrates. The configuration was the same in all the diodes: two heavily doped n-type electrodes with an undoped active region in between. The electrode next to the substrate (`emitter') was made of 500 \AA \, of n$^+$In$_{0.53}$Ga$_{0.47}$As ($1\times10^{19}$ cm$^{-3}$) and 2000 \AA \, of n$^+$ In$_{0.53}$Ga$_{0.47}$As ($1\times10^{18}$ cm$^{-3}$) followed by an undoped In$_{0.53}$Ga$_{0.47}$As spacer layer of 50 \AA. The top electrode (`collector') had the same structure as the bottom electrode. The active region of sample A consisted of the following: 100 \AA \, of In$_{0.52}$Al$_{0.48}$As (barrier), 82 \AA \, of In$_{0.53}$Ga$_{0.47}$As (well), 52 \AA \, of In$_{0.52}$Al$_{0.48}$As (barrier), 53 \AA \, of In$_{0.53}$Ga$_{0.47}$As (well), and 100 \AA \, of In$_{0.52}$Ga$_{0.48}$As (barrier). Sample B was identical to sample A except for the thickness of the middle barrier, which was 100 \AA. The active region of the DBRT (sample C) consisted of 100 \AA \, of In$_{0.52}$Al$_{0.48}$As (barrier), 53 \AA \, of In$_{0.53}$Ga$_{0.47}$As (well) and 100 \AA \, of In$_{0.52}$Al$_{0.48}$As (barrier). The diodes were defined by photolithography and wet etching to a size of 20x20 $\mu$m$^2$. 

Transport and shot noise measurements were done at 4.2K with the device immersed in liquid helium. The current-voltage characteristic of each sample was determined by biasing it through a low noise, battery-powered voltage follower (that reduced the source impedance from the voltage source) and recording the voltage drop across a calibrated resistor in series with the sample. Conductance was measured using an AC modulation voltage with 0.1mV rms amplitude and detecting the corresponding drop across that resistor with a lock-in amplifier. Positive bias is defined here as the voltage polarity for which electrons tunnel from the emitter into the wider quantum well and then into the narrower well.

To measure noise, the samples were connected in series to a very low noise, battery-powered current amplifier. The output noise of the amplifier, which consists of the noise from the sample, amplifier noise, thermal noise and background noise, was measured by a spectrum analyzer at 5kHz at 1mV intervals. The sample's noise was determined as follows: first, the noise data recorded by the spectrum analyzer were divided by the amplification gain; then, the amplifier and thermal noises were subtracted; finally, the background noise, determined using several calibrated metal film resistors, was subtracted. 

Since the results from the two TBRTDs were similar, in the following we will focus on sample A, whose conduction band profile at zero bias is shown in Fig.\ref{fig:Conduction}. The bound-state energies in the wider (narrower) well, denoted by $w_1$ ($n_1$) and $w_2$ ($n_2$), are 41 (93) meV and 235 (425) meV above the Fermi level, respectively.\cite{nonpar} Under positive bias, the energy separation between $w_1$ and $n_1$ diminishes and at a certain voltage both levels become aligned (resonant condition) while being below the Fermi level. Conservation of energy and parallel (to the layers' planes) momentum favor electron tunneling at that voltage and, as a result, the current has a strong spike. Ideally, at voltages below or above resonance the current should be negligible. The situation repeats itself when $w_2$ and $n_2$ become aligned at even higher voltage and, under reverse bias, when $w_2$ and $n_1$ are in resonance.

\begin{figure}[thp!]
  \centering
  \includegraphics[width=3.2in,angle=0,clip=]{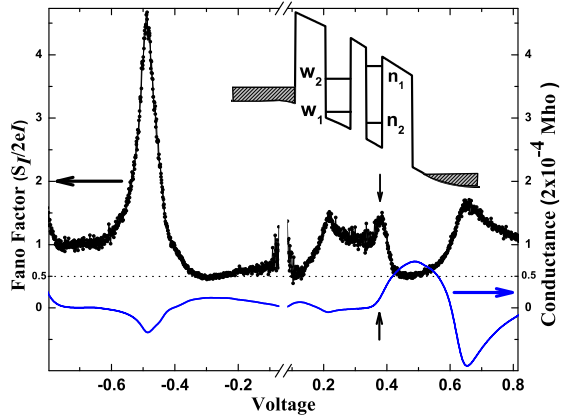}
  \caption[Fano Factor of a Triple Barrier Resonant Tunneling Diode]{\label{fig:Fano} Experimental Fano factor (circles) and conductance (solid line) of sample A as a function of voltage. Also shown is a diagram of the conduction band profile at a voltage between the $w_1 - n_1$ and $w_2 - n_2$ resonances. The vertical arrows at 0.38 V point to an anomalous peak in the noise characteristic and a corresponding (weak) feature in the conductance. The Fano factor in the region between -0.1V and 0.1V is not shown because its uncertainty was extremely large due to the very small current, in comparison with that at other voltages.} 
\end{figure}

As shown in Fig. \ref{fig:noise}, the $I$-$V$ characteristic of sample A at T = 4.2K exhibits the predicted behavior, broadly speaking. The two current peaks in forward bias correspond to the $w_1-n_1$ and $w_2-n_2$ resonances, while the peak in reverse bias is for the $n_1-w_2$ resonance. (At 77K the $I$-$V$ characteristic did not change much, but at 300K, although the $w_1-n_1$ and $n_1-w_2$ were still apparent, there was a substantial thermally activated current background.) There is, though, a clear difference between the predicted and observed behavior. Experimentally the current rise is gradual (see, e.g., the voltage region between 0.075 and 0.175 V or between -0.1 V and -0.4 V, in Fig. \ref{fig:noise}), which contrasts with the delta-function-like  characteristic that energy and momentum conservation demand. This difference is attributed to the non-conservation of parallel momentum, and will be discussed in detail elsewhere.\cite{Lin2004}

Figure \ref{fig:noise} also shows the measured shot noise characteristic for sample A and compares it with the Poissonian value $2eI$. As it is apparent in the figure, the shot noise is reduced below $2eI$ whenever the current rises and is enhanced when the current drops. This behavior is qualitatively similar to that found in sample C and in previous reports about noise in DBRTDs.\cite{Li90, Iannaccone98, Kuznetsov98, Song2003} The deviation of the shot noise from the Poissonian value is best illustrated by plotting the Fano factor $F$, shown in Fig \ref{fig:Fano} along with the conductance. For the $w_1-n_1$ peak, it is $F = 0.55\pm 0.06$ at 0.17V (the lowest voltage at which the current is sufficiently high to make a meaningful determination of noise in our set-up) and then $F$ increases gradually, passing the value of 1 and reaching a local maximum of 1.4 when the differential conductance is negative and has a minimum value (at V = 0.23V). Further on, still in the NDC region, $F$ goes back to 1 but then increases and has a new maximum at 0.38V, before decreasing and merging with the features of the $w_2-n_2$ peak. 

\begin{figure}[thp!]
  \centering
  \includegraphics[width=3.2in,angle=0,clip=]{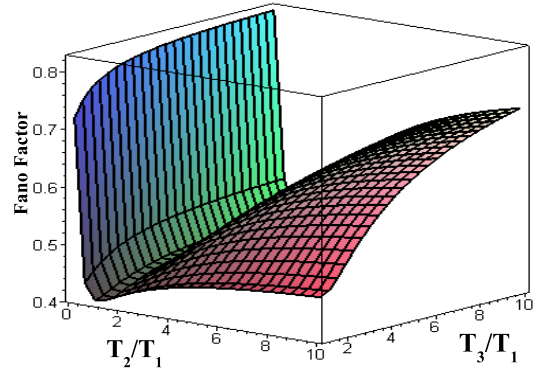}
  \caption[Current Voltage Characteristics of a Barrier Resonant Tunneling Diode]{\label{fig:sim} Simulated three-dimensional plot of the calculated Fano factor in a triple barrier structure in which the tunneling probabilities through the individual barriers are given by $T_1, T_2$, and $T_3$. The calculation was done using a semiclassical model that assumes sequential tunneling throughout the structure. The Fano factor has a minimum value of 0.41 when the tunneling probability is the same for all three barriers. When the probabilities are very different from each other the Fano factor approaches unity.
}
\end{figure}

For this second peak, $F$ increases from an initial value of $0.50\pm 0.02$, reaches a maximum value of 1.5 when the conductance is minimum (V = 0.65V), and then gradually reverts to 1. The same behavior is observed for the $n_1-w_2$ peak in reverse bias, with a minimum $F$ of $0.48\pm 0.02$ and a maximum of 4.6. There is a noticeable difference, however, between both polarities: the enhancement of the Fano factor for the current peak under reverse bias is at least three times larger than for any of the two peaks under forward bias. For comparison, in sample C (the control DBRTD) the minimum and maximum Fano factors were found to be $0.51\pm 0.02$ and $1.2\pm 0.1$, respectively. 

At this point, it is worth summarizing the experimental facts. First, the noise behavior of both TBRTDs is qualitatively similar to that of the control DBRTD and other DBRTDs studied before,\cite{Li90, Iannaccone98, Kuznetsov98, Song2003} but it contrasts with the unusual noise dependence in a TBRTD reported earlier.\cite{Yau97} Second, the minimum value of $F$ is around 0.5 for the two TBRTDs we have studied, even though the thickness of the central barrier in sample A was quite different from that of the end barriers. Third, there is an unusual (local) maximum Fano factor, without a corresponding well-defined feature in the conductance.  Fourth, the enhanced Fano factor for the $n_1-w_2$ peak is three times larger than for any of the other two peaks, even though the corresponding minimum in the conductance is much less pronounced for $n_1-w_2$ than for $w_2-n_2$.

To compare our experimental minimum $F$ values with theoretical predictions, we calculated the shot noise using a sequential-tunneling model developed to treat shot noise in a multiple-barrier system.\cite{DeJong95} Given the barrier thickness of our samples, it is appropriate to see the tunneling process as sequential, a regime in which quantum and semi-classical models give the same answer.\cite{Sun99}

The results of our calculations are summarized in Fig.\ref{fig:sim}, where in a simulated three-dimensional plot we represent the Fano factor as a function of the ratios $T_2/T_1$ and $T_3/ T_1$ ($T_i$ is the transmission probability through the $i$-th barrier, considered individually). If the tunneling probability through the emitter and collector barriers is the same, that is, $T_3=T_1$, and the central barrier is not thicker than either of the two end barriers ($T_2 \ge T_1$) then, according to Fig.\ref{fig:sim}, the minimum Fano factor should range between 0.41 (when $T_2$ =$ T_1$) and 0.5 (when $T_2 \gg T_1$, that is, in the DBRT limit). Thus the calculation predicts that at zero bias the Fano factor for sample A should be close to 0.5 while that of sample B should approach 0.41.

When a bias is applied to the TBRTD the potential profile is affected and the various tunneling probabilities can change considerably. Using a Schr\"{o}dinger-Poisson solver and the transfer matrix method\cite{Mendez87,nonpar} to calculate T(E) for each tunnel barrier, we have obtained $T_3$ and $T_1$ as a function of voltage. For sample A, at V = 0.11V (the onset of the $w_1-n_1$ peak) the probability ratio $T_3/T_1$ thus determined was 7.9 and the corresponding Fano factor was F = 0.80, to be compared with an experimental value of $0.55\pm 0.06$. For sample B (at V = 0.17V) the calculated and experimental values for F were 0.51 and $0.44\pm 0.06$, respectively.

For resonances at higher voltages, the discrepancy between calculation and experiment is larger. Thus, in sample A, at V = 0.46V (the current onset of the $w_2-n_2$ peak), $T_3/T_1 > 400$ and the calculated factor was 1.0. For V = - 0.29V, (the onset of $n_1-w_2$ peak) $T_3/T_1 \approx 50$ and $F = 0.98$. In sharp contrast, experimentally, for both voltages it is F $ \approx  0.5$. This large difference between theory and experiment, also observed in sample B, is well outside our experimental uncertainty and remains unexplained. (As a reference, for the control DBRTD that difference was minimal: 0.55 vs $0.51\pm 0.06$.)

The enhancement of the shot noise at 0.38 V, marked by vertical arrows in Fig.\ref{fig:Fano} and also observed in sample B, may be due to phonon-assisted tunneling via the emission of a LO-phonon in the InGaAs layer. (There is also a hint of a related feature in the conductance). The voltage at which the enhancement occurred is consistent with that interpretation, but it is not clear why then a similar enhancement is not observed for the other two resonances. 

The large enhancement of the Fano factor we have observed in the NDC region of the $n_1-w_2$ peak deserves special attention, since it runs counter to our understanding of the origin of noise enhancement in a DBRTD\cite{Iannaccone98, Kuznetsov98} (and by extension in a TBRTD), which is as follows. Since in the NDC region the quantum-well level is already below the conduction-band edge of the emitter, the density of states (DOS) for tunneling into the well is quite small. But when an electron does tunnel it charges the well and modifies the potential profile, pushing the center of the DOS higher in energy and thus enhancing the probability for a second electron to tunnel. This positive correlation increases the shot noise; the sharper the density of states the more pronounced is the NDC -- and the larger the shot-noise enhanced should be. 

This intuitive prediction has been confirmed experimentally in InAs-AlSb-GaSb DBRTDs, in which, using a magnetic field to control the strength of the NDC, it was found that F increased monotonically with increasingly stronger NDC.\cite {Kuznetsov98} In both of our TBRTDs the correlation holds when we compare the strengths of the NDC for the $w_1-n_1$ and the $w_2-n_2$ resonances, but it breaks down when the $n_1-w_2$ resonance is included (see Fig. 3). 

Regarding noise, no polarity asymmetry was found in the control DBRTD. What makes then the forward- and reverse-bias current peaks different in TBRTDs that could affect the enhancement of noise? The only apparent difference between the $n_1-w_2$ peak and either the $w_1-n_1$ or the $w_2-n_2$ peak lies in the relative symmetry of the quantum states involved in the tunneling process. At zero bias, the wavefunctions of the $w_1$ and $n_1$ states are symmetric, relative to the center of their corresponding wells, while the $w_2$ and $n_2$ are antisymmetric. It is unclear, though, how this different symmetry could affect shot noise, especially at high bias, when the symmetry of the wavefunction is greatly reduced. 

Some light might be shed into this unresolved puzzle by measuring the shot noise of a TBRTD identical to sample A (or sample B) but with the order of the two quantum wells reversed relative to the sample's substrate. If the anomalous noise enhancement that we have observed is indeed only a consequence of the electron wavefunctions' asymmetry then the new $I$-$V$ and noise characteristics should be the same as those in Figs. 2 and 3, but with opposite polarity. Were this the case, it would then be most interesting to explore the effect of the central barrier on that enhancement and, naturally, to approach the regime of strongly coupled wells, where coherence might also affect shot noise.

We acknowledge useful discussions with Dr. H. Grahn. The work has been sponsored by National Science Foundation under Grant No. DMR-0305384.


\begin{thebibliography}{}
\bibitem {nonpar}
We have used parabolic band structures with effective masses of  $m^*_{InGaAs}=0.047m_0$ and $m^*_{InAlAs}=0.077m_0$, where $m_0$ is the free electron mass, and a band offset of 526 meV.
\end{thebibliography}
\end{document}